%% file: 0v4bPandaX4T.tex
\documentclass[
twocolumn,
superscriptaddress,
amsmath,amssymb,
aps,
prd,
]{revtex4-2}

\usepackage{lipsum}
\usepackage{array}
\usepackage{graphicx}
\usepackage{xcolor}
\usepackage{amsmath}
\usepackage{booktabs}
\usepackage{multirow}
\usepackage{natbib}
\setcitestyle{numbers,comma,square}
\usepackage{lineno}
\usepackage{comment}

\usepackage{subfigure}
\usepackage[colorlinks=true, urlcolor=black, citecolor=blue, linkcolor=blue, allcolors=blue]{hyperref}
\setcitestyle{etalnum=3}

\usepackage{orcidlink}

\begin{document}

\title{Search for neutrinoless quadruple beta decay of $^{136}$Xe in PandaX-4T detector}
\input{authorlist}

\maketitle

\vskip 1.5mm

\onecolumngrid
The observation of neutrinoless quadruple beta decay (0$\nu$4$\beta$) in the absence of neutrinoless double beta decay (0$\nu$2$\beta$) has been argued to provide a strong indication that neutrinos are Dirac particles. 
We report a search for 0$\nu$4$\beta$ decay of $^{136}\text{Xe}$ using a total $^{136}\text{Xe}$ exposure of 148.4~kg$\cdot$yr, collected during the commissioning and the first science runs of the PandaX‑4T experiment. No significant excess of events over the background is observed. A lower limit on the 0$\nu$4$\beta$ decay half‑life of $^{136}\text{Xe}$ is set at 6.01 $\times$ 10$^{24}$~yr at the 90\% confidence level. This result establishes the most stringent constraint on this process in xenon, demonstrating the unique capability of the PandaX-4T detector in probing lepton number violation and shedding light on the fundamental nature of neutrinos.

\twocolumngrid

\section{Introduction}
The search for rare radioactive decays provides a crucial window into physics beyond the Standard Model of particle physics. The neutrinoless double beta decay (0$\nu$2$\beta$) is the flagship search for revealing the Majorana nature of neutrinos~\cite{add_FURRY_DBD,add_NLDBD_Schechter}. It could provide direct evidence for the violation of the lepton number L by 2 ($\Delta$L = 2), and constrain the absolute mass scale of neutrinos~\cite{add_theory_0v2b,add_Matteo_0v2b,add_Michelle_0v2b,add_Juan_0v2b}. There are many experiments investigating 0$\nu$2$\beta$ decay in different isotopes today, such as $^{76}$Ge~\cite{13_GERDA_NLDBD,13_MAJORANA_Ge,13_CDEX_Ge,add_LEGEND}, $^{130}$Te~\cite{add_CUORE_latest}, and $^{136}$Xe~\cite{12_EXO_NLDBD,add_KLZ_latest}, but it is still escaping detection after decades of search. 
On the other hand, the neutrinoless quadruple beta decay (0$\nu$4$\beta$) was first proposed by Heeck and Rodejohann~\cite{6_Heeck_2013} as a distinctive signature of $\Delta$L = 4 processes, which can arise even for Dirac neutrinos while forbidding the conventional $\Delta$L = 2 channels. Thus, observing 
0$\nu$4$\beta$ decay in the absence of 
0$\nu$2$\beta$ decay would provide a strong indication in favor of the Dirac nature of neutrinos, barring highly fine-tuned cancellation scenarios~\cite{7_Hirsch_Dirac}. Moreover, such 
$\Delta$L = 4 scenarios have also been linked to dark matter candidates and CP violation in the lepton sector~\cite{add_candi_DM, add_CP_viola}. Nevertheless, the experimental search for 
0$\nu$4$\beta$ decay remains extremely challenging owing to its expectedly long half-life, which is highly sensitive to the mass scale $\Lambda_{\rm NP}$ characterizing the new $\Delta$L = 4 interactions~\cite{6_Heeck_2013}.
Theoretically, there are only three isotopes that can undergo this process: $^{136}\text{Xe}$ ($\text{Q}_{4\beta}$ = 0.079~MeV), $^{96}\text{Zr}$ ($\text{Q}_{4\beta}$ = 0.642~MeV), and $^{150}\text{Nd}$ ($\text{Q}_{4\beta}$ = 2.084~MeV)~\cite{6_Heeck_2013,10_AME2016}. 
To date, there has been no experimental search for $^{96}\text{Zr}$. For $^{136}\text{Xe}$ and $^{150}\text{Nd}$, the 90\% confidence level (C.L.) lower bounds on the half‑life are 3.7 $\times$ 10$^{24}$~yr from XMASS~\cite{8_XMASS_0v4b} and (1.1-3.2) $\times$ 10$^{21}$~yr from NEMO‑3~\cite{9_NEMO3_0v4b}, respectively.

The dual-phase xenon time projection chamber (TPC) technology has seen significant achievements in direct dark matter detection over the last several decades~\cite{9_DM_pandax,9_DM_LZ,add_DM_XENONnT_latest}. Furthermore, owing to the 8.9\% natural abundance of $^{136}\text{Xe}$, the natural xenon experiments, such as XENON~\cite{15_XENON_NLDBD} and PandaX~\cite{24_NKX_NLDBD,24_ZS_NLDBD}, exhibit substantial potential to search for $^{136}\text{Xe}$ 0$\nu$2$\beta$ decay. The future multi-ten-ton scale liquid xenon (LXe) detectors will provide stringent tests on the Majorana nature of neutrinos using the 0$\nu$2$\beta$ decay of $^{136}\text{Xe}$~\cite{16_XT,add_DARWIN}, effectively covering most of the parameter space for inverted neutrino mass ordering.
This also provides us with an excellent opportunity for the search of 0$\nu$4$\beta$ decay.  In this letter, we report a search for 0$\nu$4$\beta$ decay of $^{136}\text{Xe}$ using the datasets from the commissioning run (Run0) and the first science run (Run1) of the PandaX-4T experiment with a total $^{136}\text{Xe}$ exposure of 148.4~kg$\cdot$yr in the energy region of interest (ROI) from 20~keV to 140~keV.

\section{detector}
PandaX-4T is a LXe TPC experiment situated in Hall B of the China Jinping Underground Laboratory (CJPL-II)~\cite{17_CJPL}.  A cylindrical TPC contains approximately 3.7~tonnes of LXe as its active volume. The volume is enclosed by 24~high-reflectivity PTFE panels with an inscribed diameter of 1185~mm. Three electrode grids (anode, gate, and cathode) are vertically arranged within the TPC to generate an extraction and drift electric field, with distances of 10~mm and 1185~mm, respectively. Two sets of three-inch Hamamatsu R11410-23 photomultiplier tube (PMT) arrays are positioned at the top and bottom of the TPC to capture the light signals generated within. The TPC is housed within a double-layered stainless steel (SS) vessel to maintain temperature stability. The vessel is nested inside a large SS water tank. Approximately 900 m$^3$ of ultra-pure water serves as a critical passive shield against external radioactivity. 

Nuclear recoil or electronic recoil within the LXe active volume generates two distinct signals: a prompt scintillation signal ($S1$) and ionization electrons. Driven by the drift field, these electrons migrate upward into the gas phase, where the stronger extraction field triggers electroluminescence, generating a delayed electroluminescence signal ($S2$). Energy reconstruction is performed using the integrated areas of $S1$ and $S2$. The vertical (Z) coordinate is derived from the $S2$-$S1$ time delay and the electron drift velocity, while the horizontal (X, Y) positions are reconstructed via maximum likelihood estimation of the $S2$ charge distribution pattern observed by the top PMT array. More details about the detector can be found in Ref.~\cite{19_mengyue}.

\section{data process and reconstruction}
The combined data from Run0 and Run1 are utilized in this work. The analysis pipeline, including data processing, signal reconstruction, and data selection criteria, is mainly inherited from the previous study on two-neutrino double electron capture (2$\nu$DEC) of $^{124}$Xe~\cite{20_bzh_xe124,21_signa_model}, with an extended energy ROI from [25, 75]~keV to [20, 140]~keV to capture the $^{136}\text{Xe}$ 0$\nu$4$\beta$ signal at $\text{Q}_{4\beta}=79$~keV.
The lower boundary is chosen to avoid the contribution of tritium~\cite{39_zxn_ALP}, and the upper boundary is chosen to avoid the prominent 163.9~keV peak from $^{131\text{m}}$Xe produced in the neutron calibration runs.

\begin{table*}[htpb]
    \centering
    \setlength{\tabcolsep}{8pt} 
    \caption{Summary of systematic uncertainties from the detector response model, quality cut efficiency, and signal selection.}
    \label{sys_uncer}
\begin{tabular}{cccc}
    \hline
    \noalign{\vskip 1.5pt}
    \hline    \multicolumn{2}{c}{Sources}               & \multicolumn{2}{c}{Values}      \\
    \cline{3-4}                &                        & Run0                      &  Run1 \\
    \hline    Detector response&a [$\sqrt{\text{keV}}$] &0.4258  $\pm$ 0.0149       & 0.4070 $\pm$ 0.0092    \\
                               &b                       &-0.0006 $\pm$ 0.0021       & 0.0020  $\pm$ 0.0010           \\
    \cline{2-4}                &c                       & 0.0022 $\pm$ 0.0014       & 0.0014  $\pm$ 0.0018             \\
                               &d [keV$^{-1}$]          & -1.0931e-05$\pm$1.1962e-05&-5.0260e-06  $\pm$ 8.8172e-06           \\
    \hline  Quality cut efficiency &     p0                 &  \multicolumn{2}{c}{0.9974 $\pm$ 0.0003}              \\
                               &     p1                 &  \multicolumn{2}{c}{-0.0360 $\pm$ 0.0058  }              \\
                               &     p2                 &  \multicolumn{2}{c}{-0.0820  $\pm$ 0.0135}              \\
    \hline    Signal selection &  Fiducial mass [tonne] &  2.381  $\pm$ 0.036     &2.477  $\pm$  0.049 \\
                               &$^{136}$Xe abundance    &  \multicolumn{2}{c}{(8.58  $\pm$ 0.11)\%}\\
                               &LXe density [g/cm${^3}$]&  \multicolumn{2}{c}{2.8502  $\pm$ 0.0036}\\
    \hline
    \noalign{\vskip 1.5pt}
    \hline
    \end{tabular}
    \end{table*}

\begin{figure}[hbp]  
    \centering 
    \includegraphics[scale=0.43]{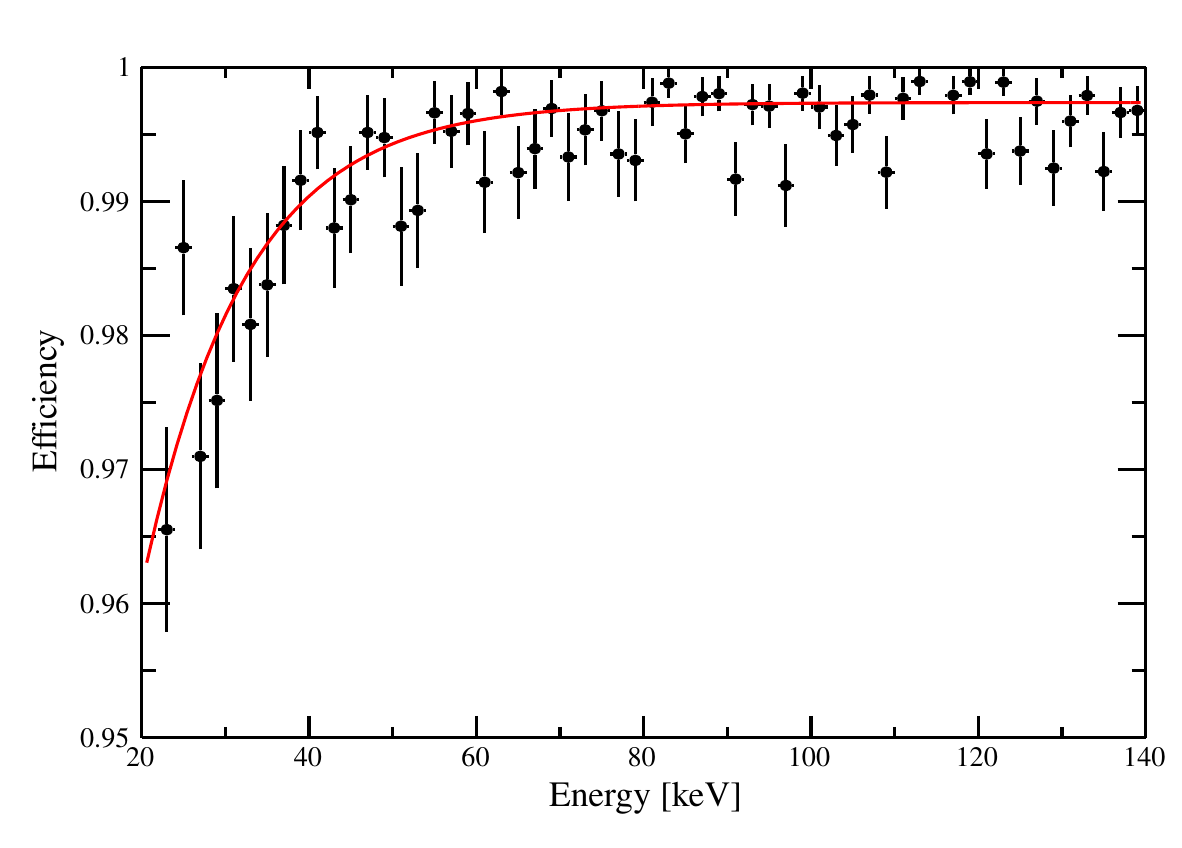}  
    \caption{The data quality cut efficiency derived from $^{220}$Rn and $^{222}$Rn calibration data (black dots). The red curve represents the exponential fit to the data.} 
    \label{fig:eff_fit}  
  \end{figure}

This analysis adopts the fiducial volume (FV) from Ref.~\cite{20_bzh_xe124}, corresponding to a total $^{136}$Xe exposure of 148.4~kg$\cdot$yr.
The $^{136}\text{Xe}$ abundance of (8.58 $\pm$ 0.11)\% is adopted from our own measurement~\cite{37_lly}.
LXe density of (2.8502 $\pm$ 0.0036)~g/cm$^{3}$ is estimated by taking into account the pressure fluctuation during detector operation~\cite{23_NME}.
The FV mass is then derived as 2.38 $\pm$ 0.04~tonnes and 2.48 $\pm$ 0.05~tonnes for Run0 and Run1, respectively~\cite{20_bzh_xe124}, as summarized in Tab.~\ref{sys_uncer}.
The efficiency of quality cuts is re-evaluated for the ROI of [20, 140]~keV using $^{220}$Rn and $^{222}$Rn calibration data.
As shown in Fig.~\ref{fig:eff_fit}, the energy dependence of the efficiency is described by the exponential function $\epsilon_{\mathrm{cut}} = p_0 + p_1 \cdot exp[p_2 \cdot (E_\text{rec} - 20)]$, with $p_0$, $p_1$, and $p_2$ as free parameters and $E_\text{rec}$ as the reconstructed energy.
The fitted efficiency curve is incorporated into the spectral fitting of signal and backgrounds, with parameters constrained by the covariance matrix.

The energy reconstruction in this analysis is based on the formula of $E_{\text{rec}} = 13.7\text{~eV} \times (S1/g1 + S2_\text{b}/g2_\text{b})$, where 13.7~eV represents the average energy required to generate a quanta in LXe.
The $S1$ and $S2$$_\text{b}$ ($S2$ signal detected exclusively by the bottom PMT array) have been corrected for spatial uniformity and temporal stability~\cite{20_bzh_xe124}. 
The parameters $g1$ and $g2_\text{b}$ are obtained separately for Run0 and Run1, by fitting the monoenergetic peaks of 41.5~keV ($^{83\text{m}}$Kr), 164~keV ($^{131\text{m}}$Xe), and 236~keV ($^{129\text{m}}$Xe + $^{127}$Xe)~\cite{20_bzh_xe124}. 
The modeling of detector energy response follows the same procedure as in Ref.~\cite{20_bzh_xe124}, with the energy resolution parametrized as $\sigma(E_{\text{rec}})/E_{\text{rec}} = a/\sqrt{E_{\text{rec}}} + b$, and the energy bias as $(E_{\text{rec}}-E_{\text{true}})/E_{\text{true}} = c + d \times E_{\text{true}}$, where $E_{\text{true}}$ and $E_{\text{rec}}$ are the true and reconstructed energies, respectively, and $\rm \sigma$ denotes the standard deviation. All the above parameters ($a, b, c, d, p_0, p_1, p_2$) are summarized in Tab.~\ref{sys_uncer} and have been incorporated into the likelihood function during the final fitting process and constrained using the corresponding covariance matrix.

Our previous analyses~\cite{22_sumperWIMP,23_NME}, which had overlapping energy ranges with this work, were optimized for the MeV scale, where the material background dominates. 
To mitigate this background, a smaller FV was therefore utilized, exploiting the self‑shielding property of LXe.
In the low‑energy ROI of this work, the material background is much less significant (discussed later), thus permitting a substantially larger FV. 
Consequently, the sensitivity of $^{136}\text{Xe}$ 0$\nu$4$\beta$ decay, characterized by $m/\sqrt{B}$, where
$m$ is the FV mass and $B$ is the expected background counts in the signal region, is improved by a factor of 2.5.

\section{Signal and Background model}
In the $^{136}$Xe $0\nu4\beta$ decay, four emitted electrons are expected to deposit their full kinetic energy within the LXe, generating a monoenergetic peak centered at 79~keV~\cite{10_AME2016}, with an energy resolution of  4.7\% for Run0 and 4.8\% for Run1 from the detector response model. In the ROI, the background contributions originate from detector materials, intrinsic radioactive xenon contaminants (e.g., $^{85}\text{Kr}$, $^{222}\text{Rn}$, and $^{220}\text{Rn}$), $^{136}\text{Xe}$ 2$\rm\nu$2$\beta$, $^{124}\text{Xe}$ 2$\nu$DEC, the activated xenon isotopes, and solar neutrinos. 
Tab.~\ref{bkg_tab} summarizes the expected background counts in the ROI, together with the best-fit values that will be detailed later.

\begin{table}[htbp]
    \centering
    \setlength{\tabcolsep}{8pt} 
    \caption{Summary of the background components and corresponding expected and best-fit counts in Run0 and Run1.}
    \label{bkg_tab}
\begin{tabular}{cccc}
    \hline
    \noalign{\vskip 1.5pt} 
    \hline  Backgrounds    &    Runs  & Expected           & Best-fit                  \\
    \hline  Material       &    0     & 357 $\pm$ 14            &352         $\pm$        10 \\
                           &    1     & 692 $\pm$ 27            &682         $\pm$        19 \\
            $^{85}$Kr      &    0     &  502 $\pm$ 261          &562         $\pm$        150    \\
                           &    1     &  1618 $\pm$ 485         &2116        $\pm$        231\\
            $^{214}$Pb     &    0     & 1781 $\pm$ 107          &1763        $\pm$        99 \\
                           &    1     & 3873 $\pm$ 232          &3876        $\pm$        207 \\
           $^{212}$Pb      &    0     & 356 $\pm$ 94            &365         $\pm$        88  \\
                           &    1     & 284 $\pm$ 50            &289         $\pm$        49   \\
$^{136}$Xe 2$\nu$2$\beta$  &    0     & 986 $\pm$ 44            &943         $\pm$        30 \\
                           &    1     & 1759 $\pm$ 79           &1684        $\pm$        53 \\
$^{124}$Xe 2$\nu$DEC       &    0    &  Float                  &255     $\pm$        29 \\
                           &    1    &  Float                  &439     $\pm$        50 \\
            $^{127}$Xe     &    0    &  Float                  &50      $\pm$        20 \\
                           &    1    &  -                      &-\\ 
            $^{133}$Xe     &    0    &  Float                  &11915   $\pm$        135\\
                           &    1    &  -                      &-\\   
            $^{125}$I      &    0    &  Float                  &159     $\pm$        41 \\
                           &    1    &  Float                  &0       $\pm$        193\\    
           Solar $\nu$     &    0    & 176 $\pm$ 18            &176     $\pm$        12\\
                           &    1    & 315 $\pm$ 32            &316     $\pm$        22\\ 
    \hline
    \noalign{\vskip 1.5pt}
    \hline
    \end{tabular}
    \end{table}

The background from detector materials originates primarily from $^{60}\text{Co}$, $^{40}\text{K}$, $^{232}\text{Th}$, and $^{238}\text{U}$ present in the flange, vessel, PMTs, and others. 
The activities of these nuclides are determined via high-purity germanium measurements~\cite{25_qzc_material} and subsequent broad-energy-spectrum fitting~\cite{26_SL_Xe136}.
Their contributions within the ROI are then evaluated with the Geant4-based BambooMC simulation framework~\cite{27_CX_bambooMC} and summed into a single background component for the analysis.
The background contribution from the stainless steel platform~\cite{37_lly}, which supports the xenon cryostat within the water shield, was found to be approximately three orders of magnitude lower than that from the detector material in the ROI and is therefore neglected here.

The $^{85}\text{Kr}$ background is evaluated using the correlated $\beta$-$\gamma$ coincidences from $\rm^{85m}\text{Rb}$, following the same procedure described in Ref.~\cite{20_bzh_xe124}. Assuming an isotopic abundance of $^{85}\text{Kr}$ of $2\times10^{-11}$~\cite{28_Radionuclides_Environment}, the krypton concentrations in Run0 and Run1 are estimated as 0.52 $\pm$ 0.27~parts per trillion (ppt) and 0.94 $\pm$ 0.28~ppt, respectively.  

The $\beta$-decay of $^{214}\text{Pb}$, a progeny of the $^{222}\text{Rn}$ decay chain~\cite{29_mwb_radon}, is the dominant background source in Run1 of this analysis. 
To determine the $^{214}\text{Pb}$ concentration, we performed a broad-energy-spectrum fit following the procedure in Ref.~\cite{23_NME} but restricted to the 200-1000~keV range to avoid overlap with our ROI. The fit yielded activities of 4.20 $\pm$ 0.22~$\rm \mu Bq/kg$ for Run0 and 5.05 $\pm$ 0.27~$\rm \mu Bq/kg$ for Run1. 
Subsequently, we estimate the $^{214}\text{Pb}$ contribution to the ROI using the ratio of events in the ROI to those in the full spectrum,  which was obtained from a dedicated $^{222}\text{Rn}$ calibration run combined with a high-precision theoretical calculation~\cite{add_theory_Pb214}.
Similarly, $\beta$-decay from $^{212}\text{Pb}$ (progeny of the $^{220}\text{Rn}$ chain) provides an additional background component. Applying the same procedure to evaluate its activity yields 0.38 $\pm$ 0.10~$\rm \mu Bq/kg$ for Run0 and 0.17 $\pm$ 0.03~$\rm \mu Bq/kg$ for Run1. For the ROI contribution, we derive it from the full decay spectrum generated by the Geant4-based BambooMC simulation, extracting the ratio of events inside the ROI. 

The $^{136}\text{Xe}$ 2$\nu2\beta$ decay background is constrained by in-situ measurements of the isotope's half-life and abundance, performed using the PandaX-4T Run0 dataset with the ROI from 440~keV to 2800~keV~\cite{26_SL_Xe136}. The collaboration has since reported an updated half-life value by combining Run0 and Run1 data~\cite{23_NME}. Because the ROI of this work overlaps with that of the updated measurement, we use the earlier value in this analysis to avoid double-fitting effects.
The $^{124}\text{Xe}$ 2$\nu$DEC decay produces monoenergetic peaks (X-rays and Auger electrons) at 64.62, 37.05, 32.98, 32.11, and 31.93~keV, as reported in our previous work~\cite{20_bzh_xe124}. Since the ROI of this analysis overlaps with that study, the $^{124}\text{Xe}$ background yield is floated in the fit.

Xenon isotopes produced by cosmogenic and neutron activation are also significant contributors to the background. The cosmogenic $^{127}\text{Xe}$ (EC decay, 36.4-day half-life) produces a 33.2~keV background signal~\cite{29_xe127}. Its activity was substantially elevated during Run0 following the injection of terrestrial xenon, but becomes negligible by Run1~\cite{20_bzh_xe124} due to its short half-life.

The isotopes $^{133}\text{Xe}$ and $^{125}\text{Xe}$ are produced via neutron capture during cosmic-ray exposure and neutron-source calibration runs, and thus are present in the PandaX-4T detector. 
The isotope $^{133}\text{Xe}$ undergoes $\beta$-decay to $^{133}\text{Cs}$ with a half-life of 5.2~days. In 98.5\% of these decays, it populates the 80.1~keV state, producing a continuous $\beta$-$\gamma$ coincidence background at this energy~\cite{30_xe133}.  Its distinctive shoulder shape can be reliably modeled with the BambooMC framework. Since the ROI fully covers this structure, it is left free in the fit. The high‑intensity neutron calibration in PandaX‑4T took place only during Run0, and the short half‑life and the long interval between runs make the $^{133}\text{Xe}$ contribution substantial there but negligible in Run1.

The isotope $^{125}\text{Xe}$ decays via electron capture with a half-life of 16.9~hours, populating $^{125}\text{I}$, which has a significantly longer half-life of 59.4~days~\cite{31_xe125}. Consequently, its daughter $^{125}\text{I}$ will persist as a background source in both Run0 and Run1. The subsequent decay of $^{125}\text{I}$ proceeds via electron capture to the 35.5~keV excited state of $^{125}\text{Te}$, with approximately 80\% of captures from the K shell, 16\% from the L shell, and 3.5\% from the M shell. The corresponding total energy depositions are 67.3~keV, 40.4~keV, and 36.5~keV, respectively~\cite{32_I125}.
All background contributions arising from activated xenon isotopes are left as free parameters in the fit, as listed in Tab.~\ref{bkg_tab}.

The background arising from the elastic scattering of solar $pp$ and $^{7}\text{Be}$ neutrinos on electrons is estimated using the event rate and continuous spectrum reported in Ref.~\cite{33_pp}. A 10\% uncertainty on the solar neutrino flux, derived from the Borexino measurement~\cite{34_pp_err}, is incorporated into this analysis.

\section{Fit method and result }
We employ the binned Poisson likelihood fitting method~\cite{24_ZS_NLDBD}. The likelihood function is constructed as:
\begin{equation}
\begin{aligned}
        L = & \prod_{r=0}^{N_{run}} \prod_{i=1}^{N_{bins}} \frac{(N_{r,i})^{N_{r,i}^{obs}}e^{-N_{r,i}}}{N_{r,i}^{obs}!}& 
        \\ & \cdot  \prod_{r=0}^{N_{run}} \prod_{j=0}^{2} \mathcal{G} (\mathcal{M}_{rj};\mathcal{M}_{rj}^0,\Sigma_{rj})
        \\ & \cdot  \prod_{k=1}^{N_{G}} G(\eta_k;0,\sigma_k),
\end{aligned}
\label{eq::likelihood}
\end{equation}
where $N_{r,i}$ and $N_{r,i}^{obs}$ $(r=0,1)$ denote the expected and observed events in the $i_\text{th}$ bin of the $r$-th Run (Run0 or Run1), respectively.  $N_{r,i}$ consists of both signal and background components. $N_i$ in both Runs is defined as:
\begin{equation}
    \begin{aligned}
        N_{i} =  (1+\eta_{s}) \cdot n_{s} \cdot S_{ i}  +  \sum_{b=1}^{N_{bkg}} (1+\eta_b) \cdot n_{b} \cdot B_{i},
    \end{aligned}
    \end{equation}
where $n_{s}$ and $n_{b}$ represent the counts of the $^{136}$Xe $0\nu4\beta$ decay signal and background components, respectively.
$S_{i}$ and $B_{i}$ denote the normalized energy spectra of the signal and background, respectively, convolved with the detector response for the $r$-th Run. 
$\eta_s$ and $\eta_b$ represent the relative uncertainties in the signal selection and the background model (see Tab.~\ref{sys_uncer}), respectively. 
The Gaussian penalty term for the detector response and quality cut efficiency in the $r$-th Run is constructed as $\mathcal{G} (\mathcal{M}_{rj};\mathcal{M}_{rj}^0,\Sigma_{rj})$, where $\mathcal{M}_{rj}^0$ represents the nominal values of the nuisance parameters—two $(j = 0, 1)$ for the detector response (energy resolution and energy bias) and one $(j = 2)$ for the quality cut efficiency—and $\Sigma_{rj}$ is the corresponding covariance matrix. The Gaussian penalty terms $G(\eta_k;0,\sigma_k)$ are used to constrain the nuisance parameters  $\eta_s$ and $\eta_b$.

\begin{figure}[htbp]  
    \centering 
    \includegraphics[scale=0.432]{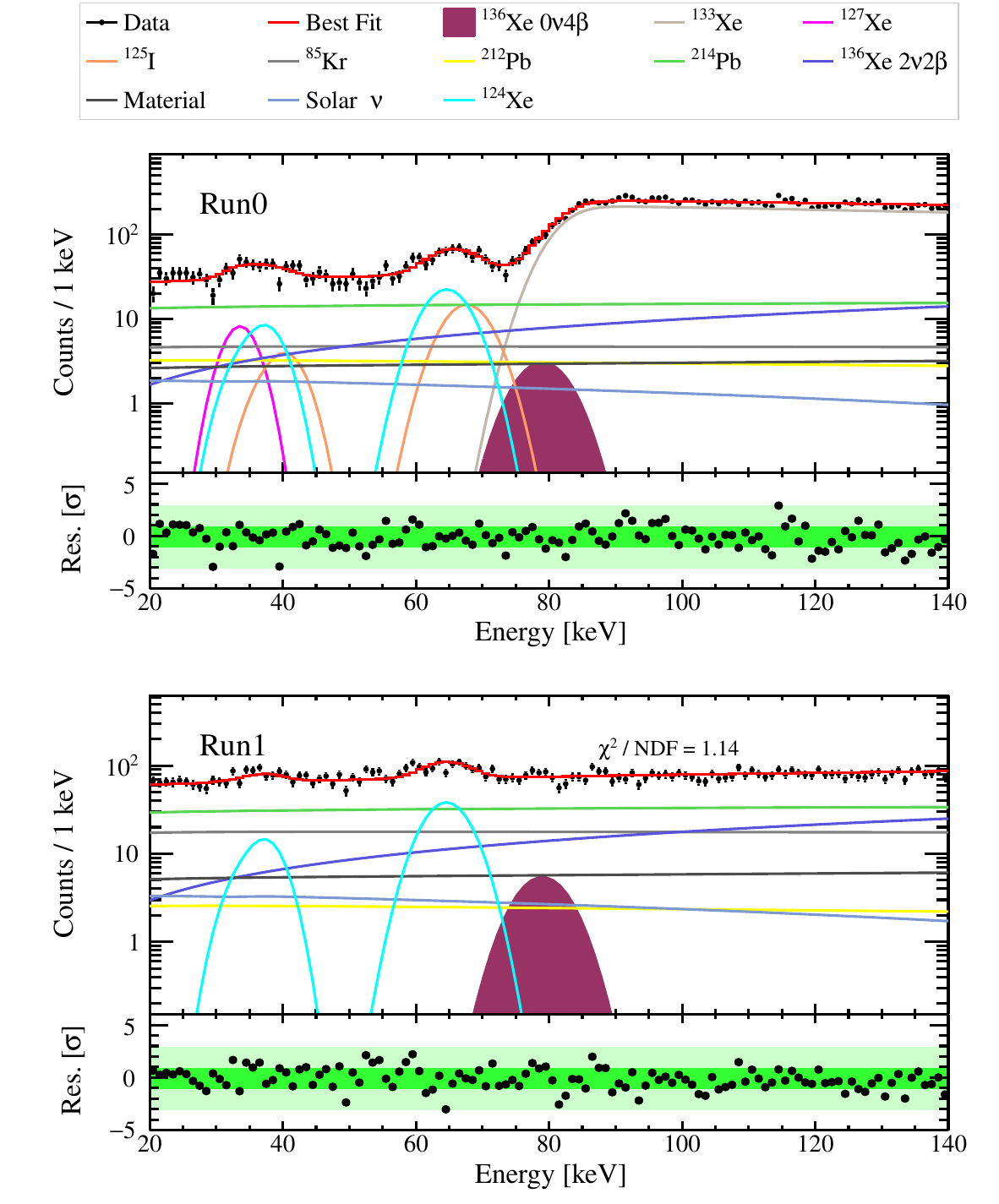}  
    \caption{Binned-likelihood fit results for Run0 (top) and Run1 (bottom) over 20-140~keV. The magenta peaks (the integral area corresponding to counts of the 90\% C.L. upper limit) mark the signal positions for better visualization. The lower panels show the residuals ($\rm (data-best\text{-}fit) / \sqrt{data}$), with dark (light) green bands representing the 
    $\pm1\sigma$ ($\pm3\sigma$) confidence regions.}
    \label{fit_result}  
  \end{figure}

\begin{figure}[t]  
    \centering 
    \hspace{-1.0cm}
    \includegraphics[scale=0.55,angle=-90]{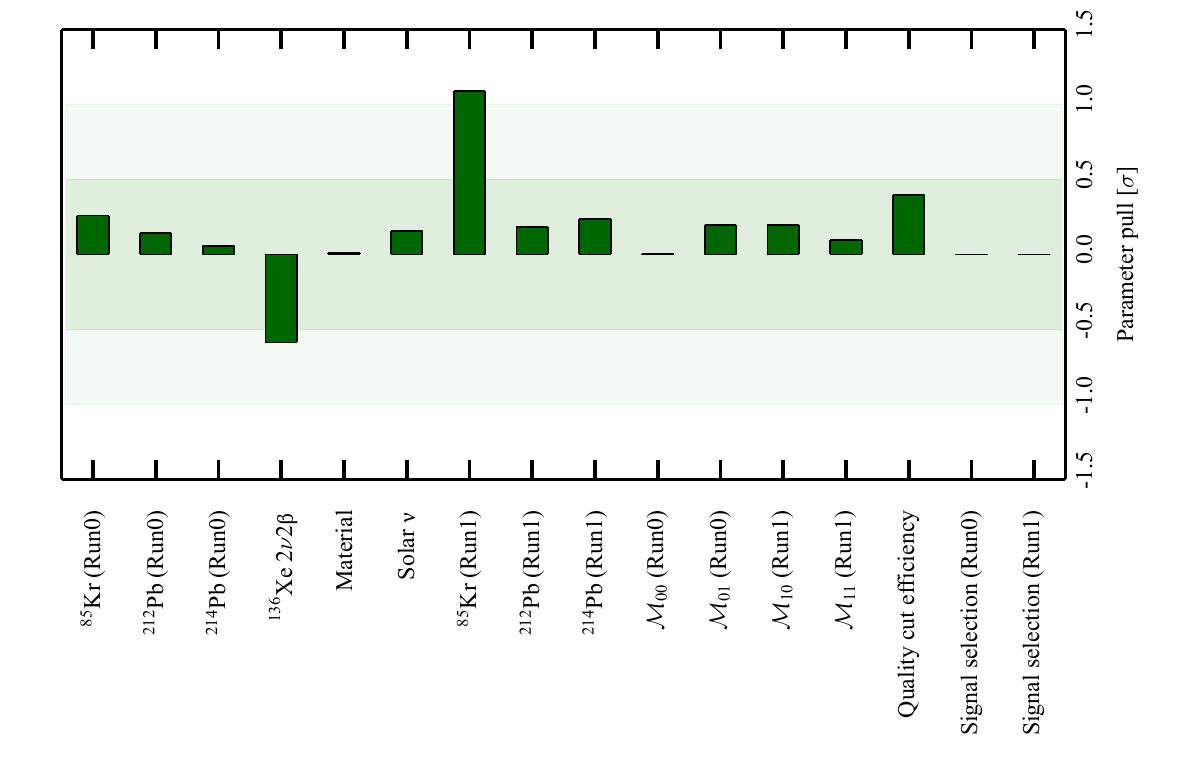}  
    \caption{Pulls of the nuisance parameters (in units of $\sigma$) from the fit. The dark (light) green bands represent the $\pm0.5\sigma$ ($\pm1\sigma$) regions.}  
    \label{pull}  
  \end{figure}

A background-only fit is first performed on the Run0 and Run1 spectra, giving a goodness‑of‑fit p‑value of 0.073.
Using the best-fit parameters from background-only fits, we generate toy Monte Carlo datasets and refit them under the signal-plus-background hypothesis. 
The median sensitivity of the lower limit on $^{136}\text{Xe}$ 0$\nu$4$\beta$ decay half-life is thus estimated as $\rm 5.5 \times 10^{24}$~yr at the 90\% C.L.. 
A subsequent fit under the signal-plus-background hypothesis is then performed, as shown in Fig.~\ref{fit_result}, giving a goodness‑of‑fit p‑value of 0.073.
No significant signal over the background is observed, and a 90\% upper limit on the signal rate is obtained as 43.75~t$^{-1}$yr$^{-1}$. 
This corresponds to a lower limit on the $^{136}\text{Xe}$ 0$\nu$4$\beta$ decay half-life of $T^{0\nu4\beta}_{1/2} > 6.01 \times 10^{24}$~yr at 90\% C.L., which is consistent with the median sensitivity within $1\sigma$.
In comparison, our previous search for axion-like particles and dark photons~\cite{22_sumperWIMP} spanned the 20-1050~keV range and already covered the signal peak of $^{136}\text{Xe}$ 0$\nu$4$\beta$ decay. 
The new half-life limit is stronger than that of the previous analysis by a factor of 3, owing to the larger FV and improved signal-to-noise ratio.

The fitted background contributions are consistent with the expected values summarized in Tab.~\ref{bkg_tab}. The dominant backgrounds are $^{133}$Xe and $^{214}$Pb for Run0 and Run1, respectively. The fitted yield of $^{124}$Xe 2$\nu$DEC, treated as a background component here, agrees with our previous analysis~\cite{20_bzh_xe124} within uncertainties. 
The pulls of all nuisance parameters under the signal-plus-background hypothesis lie within the 2$\sigma$ range, as shown in Fig.~\ref{pull}. The pull of $^{85}\text{Kr}$ of Run1 is slightly higher than the others due to its weak constraint, but remains within 1.1$\sigma$.

\section{Summary}
In summary, a binned-likelihood search for $^{136}\text{Xe}$ 0$\nu$4$\beta$ decay is conducted using the combined Run0 and Run1 data from the PandaX-4T experiment. The fit reveals no significant excess above background. At 90\% C.L., the derived lower limit on the half-life is $T^{0\nu4\beta}_{1/2} > 6.01 \times 10^{24}$~yr, which is improved by a factor of approximately 1.6 over the previous XMASS result~\cite{8_XMASS_0v4b} and constitutes the most stringent limit on this process from xenon detectors to date. 

With larger statistics and improved detector performance, the Run2 data~\cite{38_pp_run2} are being analyzed and will further enhance the sensitivity of the $^{136} \text{Xe}$ 0$\nu$4$\beta$ decay search. The upcoming PandaX-xT~\cite{16_XT} is expected to significantly improve sensitivity for both 0$\nu$2$\beta$ decay and 0$\nu$4$\beta$ decay modes. Concerning the neutrino nature, the observation of 0$\nu$2$\beta$ decay would support the Majorana hypothesis, whereas a first observation of 0$\nu$4$\beta$ decay in the absence of 0$\nu$2$\beta$ decay, barring highly fine-tuned cancellation scenarios~\cite{7_Hirsch_Dirac}, would favor the Dirac scenario.

\section{Acknowledgements}
\input{acknowledgement}

\twocolumngrid
\bibliographystyle{unsrtnat}
\bibliography{0v4bPandaX4T} 

\end{document}

%% file: authorlist.tex

\def\tdli{State Key Laboratory of Dark Matter Physics, Key Laboratory for Particle Astrophysics and Cosmology (MoE), Shanghai Key Laboratory for Particle Physics and Cosmology, New Cornerstone Science Laboratory, Tsung-Dao Lee Institute, Shanghai Jiao Tong University, Shanghai 201210, China}
\def\sjtuphys{State Key Laboratory of Dark Matter Physics, Key Laboratory for Particle Astrophysics and Cosmology (MoE), Shanghai Key Laboratory for Particle Physics and Cosmology, School of Physics and Astronomy, Shanghai Jiao Tong University, Shanghai 200240, China}
\def\MESJTU{School of Mechanical Engineering, Shanghai Jiao Tong University, Shanghai 200240, China}
\def\SPEIT{SJTU Paris Elite Institute of Technology, Shanghai Jiao Tong University, Shanghai 200240, China}
\def\SJTUSC{Shanghai Jiao Tong University Sichuan Research Institute, Chengdu 610213, China}

\def\BUAA{School of Physics, Beihang University, Beijing 102206, China}
\def\BUAACenter{Peng Huanwu Collaborative Center for Research and Education, Beihang University, Beijing 100191, China}
\def\BUAALab{International Research Center for Nuclei and Particles in the Cosmos \& Beijing Key Laboratory of Advanced Nuclear Materials and Physics, Beihang University, Beijing 100191, China}
\def\SCNT{Southern Center for Nuclear-Science Theory (SCNT), Institute of Modern Physics, Chinese Academy of Sciences, Huizhou 516000, China}

\def\USTClab{State Key Laboratory of Particle Detection and Electronics, University of Science and Technology of China, Hefei 230026, China}
\def\USTCdep{Department of Modern Physics, University of Science and Technology of China, Hefei 230026, China}

\def\YaLongSD{Yalong River Hydropower Development Company, Ltd., 288 Shuanglin Road, Chengdu 610051, China}
\def\scKeyLab{Jinping Deep Underground Frontier Science and Dark Matter Key Laboratory of Sichuan Province, Liangshan 615000, China}

\def\pku{School of Physics, Peking University, Beijing 100871, China}
\def\CHEPpku{Center for High Energy Physics, Peking University, Beijing 100871, China}

\def\SDUdep{Research Center for Particle Science and Technology, Institute of Frontier and Interdisciplinary Science, Shandong University, Qingdao 266237, China}
\def\SDUlab{Key Laboratory of Particle Physics and Particle Irradiation of the Ministry of Education, Shandong University, Qingdao 266237, China}

\def\SYU{School of Physics, Sun Yat-sen University, Guangzhou 510275, China}
\def\SYUSFI{Sino-French Institute of Nuclear Engineering and Technology, Sun Yat-sen University, Zhuhai 519082, China}
\def\SYUzhuhai{School of Physics and Astronomy, Sun Yat-sen University, Zhuhai 519082, China}
\def\SYUshenzhen{School of Science, Sun Yat-sen University, Shenzhen 518107, China}

\def\NKU{School of Physics, Nankai University, Tianjin 300071, China}
\def\YTU{Department of Physics, Yantai University, Yantai 264005, China}
\def\FDU{Key Laboratory of Nuclear Physics and Ion-beam Application (MOE), Institute of Modern Physics, Fudan University, Shanghai 200433, China}
\def\CDUT{College of Nuclear Technology and Automation Engineering, Chengdu University of Technology, Chengdu 610059, China}

\affiliation{\tdli}
\author{Zhiyuan Li}\affiliation{\SYUSFI}
\author{Peiyuan Chen\orcidlink{0009-0008-8703-4495}}\affiliation{\sjtuphys}
\author{Wei Chen\orcidlink{0009-0009-5911-7135}}\affiliation{\sjtuphys}
\author{Xiaohua Chen}\affiliation{\tdli}
\author{Xun Chen\orcidlink{0000-0001-7961-7908}}\affiliation{\tdli}\affiliation{\SJTUSC}\affiliation{\scKeyLab}
\author{Yunhua Chen}\affiliation{\YaLongSD}\affiliation{\scKeyLab}
\author{Chen Cheng\orcidlink{0000-0003-0164-7538}}\affiliation{\BUAA}
\author{Xiangyi Cui}\affiliation{\tdli}
\author{Yuxin Cui}\affiliation{\BUAA}
\author{Manna Deng}\affiliation{\SYUSFI}
% ...
\collaboration{PandaX Collaboration}
\noaffiliation

%% file: acknowledgement.tex

 

This project is supported in part by grants from National Key R\&D Program of China (Nos. 2023YFA1606200, 2023YFA1606202), National Science Foundation of China (Nos. 12305121, U23B2070, U25B2093), and by Office of Science and Technology, Shanghai Municipal Government (grant Nos. 21TQ1400218, 22JC1410100, 23JC1410200, ZJ2023-ZD-003). We thank for the support by the Fundamental Research Funds for the Central Universities. We also thank the sponsorship from the Chinese Academy of Sciences Center for Excellence in Particle Physics (CCEPP), Thomas and Linda Lau Family Foundation, New Cornerstone Science Foundation, Tencent Foundation in China, and Yangyang Development Fund. Finally, we thank the CJPL administration and the Yalong River Hydropower Development Company Ltd. for indispensable logistical support and other help. 